\RequirePackage{fix-cm}
\documentclass[twocolumn,epjc3]{svjour3}  
\smartqed  
\RequirePackage{graphicx}
\RequirePackage{amsmath}
\RequirePackage{amssymb}
\journalname{Eur. Phys. J. A}
\begin{document}

\title{Polarization transfer in $\vec{e}^+p \rightarrow e^+ \vec{p}$ scattering using the Super BigBite Spectrometer
}


\author{A.~J.~R.~Puckett\thanksref{e1,addr1}
        \and
        J.~C.~Bernauer\thanksref{addr2,addr3} 
        \and
        A.~Schmidt\thanksref{addr4}
}

\thankstext{e1}{e-mail: andrew.puckett@uconn.edu}


\institute{University of Connecticut, Storrs, CT 06269, USA \label{addr1}
           \and
           Stony Brook University, Stony Brook, NY 11789, USA \label{addr2}
           \and
           RIKEN-Brookhaven Research Center, Brookhaven National Lab, Upton, NY 11973, USA \label{addr3}
           \and
           George Washington University, Washington, DC 20052, USA \label{addr4}
}

\date{Received: date / Accepted: date}

\maketitle

\begin{abstract}
The effects of multi-photon-exchange and other higher-order QED corrections on elastic electron-proton scattering have been a subject of high experimental and theoretical interest since the polarization transfer measurements of the proton electromagnetic form factor ratio $G_E^p/G_M^p$ at large momentum transfer $Q^2$ conclusively established the strong decrease of this ratio with $Q^2$ for $Q^2 \gtrsim 1$ GeV$^2$.  
This result is incompatible with previous extractions of this quantity from cross section measurements using the Rosenbluth Separation technique. 
Much experimental attention has been focused on extracting the two-photon exchange (TPE) effect through the unpolarized $e^+p/e^-p$ cross section ratio, but polarization transfer in polarized elastic scattering can also reveal evidence of hard two-photon exchange.
Furthermore, it has a different sensitivity to the generalized TPE form factors, meaning that measurements provide new information that cannot be gleaned from unpolarized scattering alone. 
Both $\epsilon$-dependence of polarization transfer at fixed $Q^2$, and deviations between electron-proton and positron-proton scattering are key signatures of hard TPE.
A polarized positron beam at Jefferson Lab would present a unique opportunity to make the first measurement of positron polarization transfer, and comparison with electron-scattering data would place valuable constraints on hard TPE.
Here, we propose a measurement program in Hall A that combines the Super BigBite Spectrometer for measuring recoil proton polarization, with a non-magnetic calorimetric detector for triggering on elastically scattered positrons.
Though the reduced beam current of the positron beam will restrict the kinematic reach, this measurement will have very small systematic uncertainties, making it a clean probe of TPE.
\end{abstract}
%
%
\section{Introduction}
\label{sec:intro}
The discrepancy between the ratio $\mu_p G_E / G_M$ of the the proton's electromagnetic form factors extracted from polarization
asymmetry measurements, and the ratio extracted from unpolarized cross section measurements, leaves the field of form factor physics
in an uncomfortable state (see \cite{Afanasev:2017gsk} for a recent review). 
On the one hand, there is a consistent and viable hypothesis that the discrepancy is caused by non-negligible hard two-photon exchange (TPE) \cite{Guichon:2003qm,Blunden:2003sp}, the one radiative correction omitted from the standard radiative correction prescriptions \cite{Mo:1968cg,Maximon:2000hm}.
On the other hand, three recent measurements of hard TPE (at VEPP-3, at CLAS, and with OLYMPUS) found that the effect of TPE is small in the region of $Q^2 < 2$~GeV$^2/c^2$ \cite{Rachek:2014fam,Adikaram:2014ykv,Rimal:2016toz,Henderson:2016dea}.
The TPE hypothesis is still viable; it is possible that hard TPE contributes more substantially at higher momentum transfers, and can fully resolve the form factor discrepancy. 
But the lack of a definitive conclusion from this recent set of measurements is an indication that alternative approaches are needed to illuminate the situation, and it may be prudent to concentrate experimental effort on constraining and validating model-dependent theoretical calculations of TPE.
There are multiple theoretical approaches, with different assumptions and different regimes of validity~\cite{Chen:2004tw,Afanasev:2005mp,Tomalak:2014sva,Blunden:2017nby,Kuraev:2007dn}.
If new experimental data could validate and solidify confidence in one or more theoretical approaches, then hard TPE could be treated in the future like any of the other standard radiative corrections, i.e., a correction that is calculated, applied, and trusted.

VEPP-3, CLAS, and OLYMPUS all looked for hard TPE through measurements of the $e^+p$ to $e^-p$ elastic scattering cross section ratio.
After applying radiative corrections, any deviation in this ratio from unity indicates a contribution from hard TPE.
However, this is not the only experimental signature one could use. Hard TPE can also appear in a number of polarization asymmetries.
Having constraints from many orthogonal directions, i.e., from both cross section ratios and various polarization asymmetries, would be valuable for testing and validating theories of hard TPE.
As with unpolarized cross sections, seeing an opposite effect for electrons and positrons is a clear signature of TPE.

Here, we propose one such polarization measurement that could both be feasibly accomplished with a positron beam at Jefferson Lab, and contribute new information about two photon exchange that could be used to constrain theoretical models.
We propose to measure the polarization transfer (PT) from a polarized positron beam scattering elastically from a proton target, for which no data currently exist.
The proposed experiment uses the Hall A Super Big-Bite Spectrometer (SBS) to measure the polarization of recoiling protons and a lead-glass calorimeter for detecting scattered positrons in coincidence.
In the following sections, we review polarization transfer, sketch the proposed measurement, and discuss possible systematic uncertainties.

\section{Polarization Transfer}
\label{sec:pt}
In the Born approximation (i.e.~one-photon exchange), the polarization transferred from a polarized lepton to the recoiling proton is
\begin{eqnarray}
    P_t &= -hP_e\sqrt{\frac{2\epsilon (1-\epsilon)}{\tau}} \frac{G_E G_M}{G_M^2 + \frac{\epsilon}{\tau}G_E^2}, \nonumber \\
    P_l &= hP_e\sqrt{1 - \epsilon^2} \frac{G_M^2}{G_M^2 + \frac{\epsilon}{\tau}G_E^2}, \label{PTeqns}
\end{eqnarray}
where $P_t$ is the polarization transverse to the momentum transfer 3-vector (in the reaction plane), $P_l$ is the longitudinal polarization, $P_e$ is the initial lepton polarization, $h$ is the lepton helicity, $\tau \equiv \frac{Q^2}{4M^2}$ is the dimensionless 4-momentum transfer squared, $\epsilon \equiv \left[1+2(1+\tau)\tan^2 \left(\frac{\theta_e}{2}\right)\right]^{-1}$, with $\theta_e$ the electron scattering angle in the nucleon rest (laboratory) frame, is the virtual photon polarization parameter, and $G_E$ and $G_M$ are the proton's electromagnetic form factors.
The strength of the polarization transfer technique is to measure $P_t/P_l$, thereby cancelling some systematics associated with polarimetry, and isolating the ratio of the proton's form factors:
\begin{equation}
\frac{P_t}{P_l}= -\sqrt{\frac{2\epsilon}{\tau(1+\epsilon)}}\frac{G_E}{G_M}. \label{ratio}
\end{equation}
This technique has several advantages over the traditional Rosenbluth separation technique for determining form factors.
This polarization ratio can be measured at a single kinematic setting, avoiding the systematics associated with comparing data taken from different spectrometer settings.
This technique allows the relative sign of the form factors to be determined, rather than simply their magnitudes.
And furthermore, whereas the sensitivity of Rosenbluth separation to $G_E^2$ diminishes at large momentum transfer, polarization transfer retains sensitivity to $G_E$ even when $Q^2$ becomes large.
When used in combination at high $Q^2$, Rosenbluth separation can determine $G_M^2$, while polarization transfer can determine $G_E/G_M$, allowing the form factors to be separately determined.

Polarization transfer using electron scattering has been used extensively to map out the proton's form factor ratio over a wide range of $Q^2$, with experiments conducted at MIT Bates~\cite{Milbrath:1997de}, Mainz~\cite{Pospischil:2001pp}, and Jefferson Lab~\cite{Gayou:2001qt,MacLachlan:2006vw,Ron:2011rd,Paolone:2010qc,Zhan:2011ji}, including three experiments, GEp-I~\cite{Jones:1999rz,Punjabi:2005wq}, GEp-II~\cite{Gayou:2001qd,Puckett:2011xg}, and GEp-III~\cite{Puckett:2010ac,Puckett:2017flj} that pushed to high momentum transfer.
Another experiment, GEp-2$\gamma$, looked for hints of TPE in the $\epsilon$-dependence in polarization transfer \cite{Meziane:2010xc,Puckett:2011xg}.
Two other experiments made equivalent measurements by polarizing the proton target instead of measuring recoil polarization \cite{Jones:2006kf,Crawford:2006rz}.

While polarization transfer is less sensitive to the effects of hard TPE, it is not immune. Following the formalism of Ref.~\cite{Carlson:2007sp}, one finds that
\begin{eqnarray}
\frac{P_t}{P_l} &=& -\sqrt{\frac{2\epsilon}{\tau (1+\epsilon)}}\frac{G_E}{G_M} \times \Bigg[ 1 \pm
                    \textrm{Re}\left(\frac{\delta \widetilde{G}_M}{G_M}\right)  \nonumber \\
                & & \pm \frac{1}{G_E}\textrm{Re}\left(\delta\widetilde{G}_E + \frac{\nu}{M^2}\widetilde{F}_3\right) \nonumber \\
                & & \mp \frac{2}{G_M}\textrm{Re}\left(\delta\widetilde{G}_M + \frac{\epsilon\nu}{(1+\epsilon)M^2} \widetilde{F}_3\right) + \mathcal{O}(\alpha^2) \Bigg], \label{eq3}
\end{eqnarray}
with $\nu \equiv (p_e + p_{e'})_\mu (p_p + p_{p'})^\mu$, and where $\delta \widetilde{G}_E$,  $\delta \widetilde{G}_M$, and $\delta \widetilde{F}_3$ are additional form factors that become non-zero when moving beyond the one-photon exchange approximation and, crucially, depend on both $Q^2$ and $\epsilon$, whereas the one-photon-exchange form factors depend only on $Q^2$. The correction terms $\delta \widetilde{G}_E$, $\delta\widetilde{G}_M$, and $\widetilde{F}_3$ are $\mathcal{O}(\alpha)$ relative to the one-photon-exchange form factors $G_E, G_M$. The $\pm$/$\mp$ symbols in Eq.~\eqref{eq3} indicate the sign with which the two-photon-exchange amplitudes enter the observable $P_t/P_l$ depending on the lepton charge, with the upper (lower) symbol indicating the appropriate sign for $e^-(e^+)$ beams.
This particular dependence on new form factors is slightly different than what one finds when taking a positron to electron cross section ratio:
\begin{eqnarray}
\frac{\sigma_{e^+p}}{\sigma_{e^-p}} &=& 1 + 4G_M\textrm{Re}\left(\delta \widetilde{G}_M + \frac{\epsilon \nu}{M^2}\widetilde{F}_3\right) \nonumber \\
  & & - \frac{4\epsilon}{\tau}G_E\textrm{Re}\left(\delta \widetilde{G}_E + \frac{\nu}{M^2} \widetilde{F}_3\right) + \mathcal{O}(\alpha^2).
\end{eqnarray}
A measurement of the difference in polarization transfer between electron and positron scattering therefore adds information about TPE in addition to what can be learned from cross section ratios alone.

The GEp-2$\gamma$ experiment looked for the effects of TPE in polarization transfer by making measurements at three kinematic points with varying values of $\epsilon$, but with $Q^2$ fixed at 2.5~GeV$^2/c^2$~\cite{Meziane:2010xc}.
Since in the absence of hard TPE the ratio $G_E/G_M$ has no $\epsilon$-dependence, any variation with $\epsilon$ is a sign of hard TPE.
The GEp-2$\gamma$ measurement was statistically consistent with no $\epsilon$-dependence, though its measurement of the relative variation with $\epsilon$ of the longitudinal component $P_\ell$ showed deviations from the one-photon exchange expectation at the level of 1.4\%, with a statistical significance of roughly 2$\sigma$.

A measurement with positron scattering for several $Q^2$ values where the discrepancy between cross section and polarization data is large, and where the $e^-p$ polarization transfer observables have already been measured precisely, will be useful for constraining TPE effects, because deviations from the Born approximation should have the opposite sign from those in electron scattering. This will help determine if deviations are truly caused by TPE, or if they arise from systematic effects.
As the largest systematic uncertainties in polarization transfer are associated with proton polarimetry, a measurement with positrons would have largely the same systematics as an experiment with electrons.

\section{Proposed Measurement}

We propose a first set of measurements of $e^+p$ polarization transfer observables in the $Q^2$ region where the Rosenbluth-polarization discrepancy is large, using the newly constructed Super BigBite Spectrometer (SBS), which was designed to measure $G_E^p/G_M^p$ to $Q^2 \approx 12\text{ GeV}^2$ using the polarization transfer technique. Details of the planned high-$Q^2$ measurements of $\vec{e}^- p$ polarization transfer, hereafter referred to as the SBS GEP experiment, can be found in the original and jeopardy proposals to Jefferson Lab's Program Advisory Committee \cite{GEP5,GEP5_PAC47}. Despite the lower expected figure-of-merit $P^2 I$ of polarized positron beams compared to polarized electron beams, these measurements can be accomplished in a reasonable time frame owing to the large solid-angle acceptance of the new SBS apparatus.
\begin{figure*}[ht]
  \begin{center}
    \includegraphics[width=0.75\textwidth]{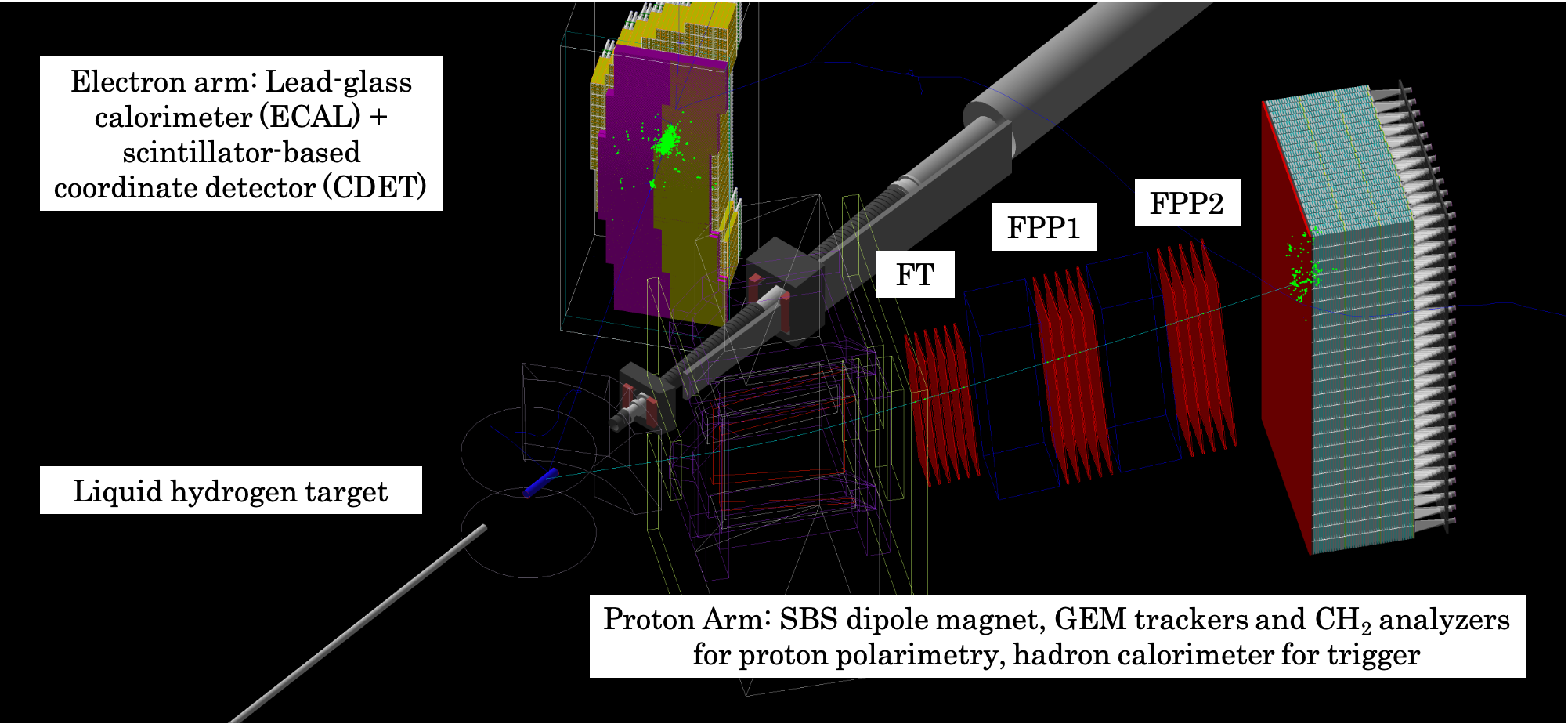}
  \end{center}
  \caption{\label{fig:SBS} Screenshot from the GEANT4-based Monte Carlo simulation of the SBS-GEP apparatus, illustrating one elastic $e^+ p$ event generated within the 40-cm liquid hydrogen target, with the electron detected in the lead-glass calorimeter (located on beam left) and the outgoing polarized proton detected in the SBS on beam right.}
\end{figure*}
Figure~\ref{fig:SBS} shows the layout of the proposed experiment in \textit{g4sbs}, the SBS GEANT4-based Monte Carlo simulation package. Polarized positrons are elastically scattered from free protons at rest in a 40-cm liquid hydrogen target. Scattered positrons are detected in a lead-glass calorimeter (ECAL) and a ``coordinate detector'' (CDET), consisting of two planes of scintillator strips with high segmentation in the vertical direction. The combination of CDET and ECAL provides a highly efficient and selective trigger for elastically scattered positrons and precise measurement of the positron's scattering angles, for a clean selection of the elastic $e^+p$ channel in the presence of higher-rate inelastic background processes, predominantly $\pi^0$ photoproduction.

Elastically scattered protons are detected in the SBS, which consists of a large dipole magnet with a transverse field integral along the direction of particle motion of up to 2.5 T$\cdot$m, a proton polarimeter with Gas Electron Multiplier (GEM)-based tracking and CH$_2$ as analyzer material, and a large hadron calorimeter. The role of the dipole magnet is for momentum analysis and to precess the longitudinal polarization of the recoiling proton into a transverse component that can be measured by the secondary analyzing scattering in the CH$_2$. The tracking in SBS relies on the relatively recently invented technology of Gas Electron Multipliers (GEMs) ~\cite{Gnanvo:2014hpa}, which can operate with stable gain at very high charged particle fluxes. The SBS front tracker, made of six GEM layers of area $40 \times 150 \text{ cm}^2$, is used for reconstruction of the proton's momentum, scattering angles, and interaction vertex, and also to define the proton's incident trajectory on the polarimeter, for subsequent measurement of the angular distribution of the secondary scattering. The spin-orbit coupling in the $\vec{p} + \text{CH}_2 \rightarrow \vec{p} + X$ scattering gives rise to an azimuthal asymmetry in the distribution of scattered protons that is proportional to their initial transverse polarization. Each of the two CH$_2$ analyzer blocks has a thickness of approximately one nuclear interaction length, and is followed by a tracker assembled from five GEM layers of area $60 \times 200\text{ cm}^2$, to measure the angular distribution of the polarization-analyzing scattering. Finally, a large iron-scintillator sampling hadronic calorimeter (HCAL), absorbs the energy of the protons and provides for efficient triggering on the events of interest, which are those in which the proton undergoes forward-angle elastic scattering in either (or both) of the two analyzers~\cite{Basilev:2019sno}.

\begin{table*}
  \caption{\label{tab:prop} Summary of proposed measurements. $E_{e}$ is the incident lepton energy, $\left<Q^2\right>$ is the acceptance averaged $Q^2$, $\theta_{e}$ is the central lepton scattering angle, $\left<\epsilon\right>$ is the acceptance averaged $\epsilon$ value, $\theta_p$ is the central proton scattering angle, and $p_p$ is the central proton momentum. The expected event rate is based on the assumption of a 200 nA (30 $\mu$A) positron (electron) beam, and $\Delta R$ is the projected absolute statistical uncertainty for the indicated number of beam days in the ratio $R \equiv -\mu_p \frac{P_t}{P_\ell} \sqrt{\frac{\tau(1+\epsilon)}{2\epsilon}}$, which equals $\mu_p G_E^p/G_M^p$ in the one-photon approximation, assuming 60\% (85\%) positron (electron) polarization. On the third line, we depict an ancillary $e^- p$ measurement at kinematics identical to the higher $Q^2\ e^+p$ measurement, that could achieve 1\% statistical precision in 24 hours (not including any time required to change CEBAF from $e^+$ to $e^-$ running).}
  \begin{center}
    \begin{tabular}{cccccccccc}
    \hline\hline\noalign{\smallskip}
      Lepton & $E_{e}$ & $\left<Q^2\right>$ & $\theta_{e}$ & $\left<\epsilon\right>$ & $\theta_p$ & $p_p$ & Event rate & Days & $\Delta R$  \\
      & GeV & GeV$^2$ & deg. & & deg. & GeV & Hz & & (absolute) \\ 
      \noalign{\smallskip}\hline\noalign{\smallskip}
      $e^+$ & 2.2 & 2.5 & 69.8 & 0.39 & 23.2 & 2.04 & 11 & 30 & 0.015 \\
      $e^{+}$ & 4.4  & 2.6 & 27.0 & 0.84 & 36.2 & 2.15 & 16 & 30 & 0.021 \\
      $e^{+}$ & 4.4  & 3.4 & 32.5 & 0.76 & 31.1 & 2.56 & 7 & 60 & 0.023 \\ \noalign{\smallskip}\hline
      $e^{-}$ & 4.4 & 3.4 & 32.5 & 0.76 & 31.1 & 2.56 & 1,050 & 1 & 0.01 \\ \noalign{\smallskip}\hline\hline
    \end{tabular}
  \end{center}
\end{table*}

Optimizing a measurement of $R_{e^+p} \equiv -\mu_p \frac{P_t}{P_\ell}\sqrt{\frac{\tau(1+\epsilon)}{2\epsilon}}$ (which equals $\mu_p G_E^p/G_M^p$ in the one-photon approximation) requires a choice of $Q^2$ and beam energy that maximizes the product of the asymmetry magnitude squared and the event rate. Merely maximizing the electron differential cross section $d\sigma/d\Omega_e$ by choosing the highest available beam energy does not always lead to the highest figure of merit (FOM) at a given $Q^2$, due to the $\epsilon$ dependence of $P_t$ and $P_\ell$, which both vanish in the limit $\epsilon \rightarrow 1$ (see equations~\eqref{PTeqns}), and also the diminishing reaction Jacobian at forward angles of the electron, where the solid angle $\Delta \Omega_e$ corresponding to the fixed proton solid angle $\Delta \Omega_p$ becomes small. The uncertainty of the ratio $R$ is typically dominated by the uncertainty of the transverse component $P_t$ of the transferred polarization, which reaches a maximum at $\epsilon \approx 0.5$, which usually occurs around $\theta_e \approx 45^\circ$. Generally speaking, for a fixed proton solid angle acceptance $\Delta \Omega_p$, the FOM at a constant $Q^2$ has a broad central maximum in the region $0.3 \le \epsilon \le 0.9$, in which it does not vary strongly. 

In planning an optimized program of exploratory $R_{e^+p}$ measurements, $Q^2$ should be chosen large enough that significant TPE corrections to this observable might reasonably be expected, but small enough that useful precision can be achieved in a ``reasonable'' amount of beam time. For an initial, exploratory set of measurements, a reasonable precision goal is $\approx 2\%$ absolute statistical uncertainty in $R_{e^+p}$ at each kinematic point (or less). Such precision would be sufficient to set a useful upper limit on the size of TPE corrections in this observable and sensitive enough to detect any TPE signal as large as, e.g., the correction needed to account for the discrepancy between cross section and polarization data. Indeed, the TPE correction to the reduced cross section (at $\epsilon = 0$ relative to $\epsilon = 1$) for elastic $ep$ scattering required to explain the discrepancy with polarization data is estimated at $\approx 4\%$ for $Q^2 \gtrsim 2$ GeV$^2$, depending on the assumed $Q^2$ and/or $\epsilon$ dependence of the TPE effect~\cite{Christy:2021snt}. 

\begin{figure}
    \centering
    \includegraphics[width=0.98\columnwidth]{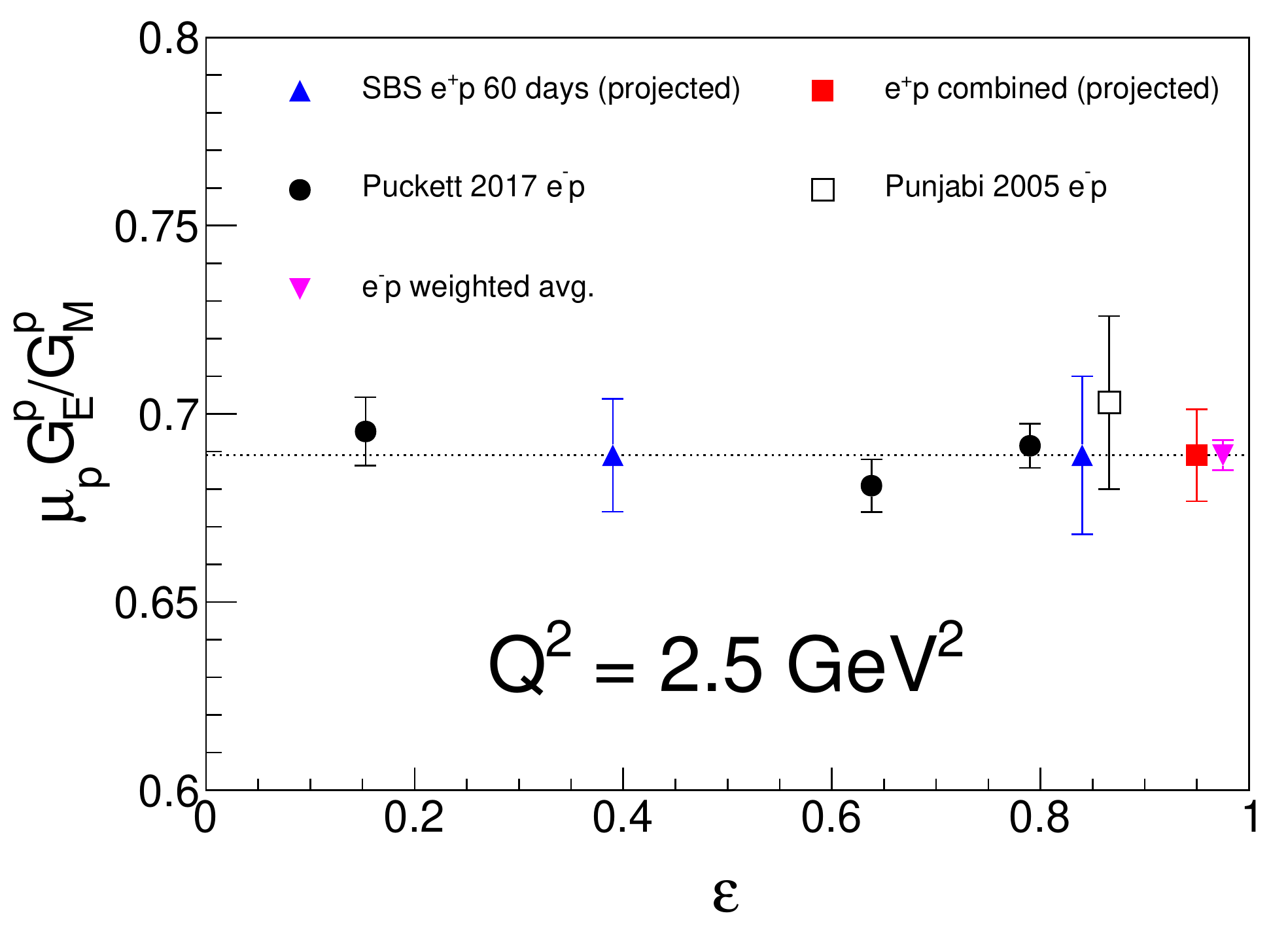}
    \caption{Projected statistical precision of a measurement of the $\epsilon$ dependence of $R(e^+p)$ at $Q^2 = 2.5$ GeV$^2$ in a hypothetical 60-day run using the SBS. The black filled circles show the existing $e^- p$ data from the Hall C GEp-2$\gamma$ experiment~\cite{Puckett:2017flj,Meziane:2010xc}. The empty square shows the GEp-I result~\cite{Punjabi:2005wq} for $R(e^-p)$ at this $Q^2$, offset by +0.03 in $\epsilon$ for clarity. The blue triangles show the projected statistical precision of the two 30-day SBS measurements depicted in Tab.~\ref{tab:prop}, plotted at the weighted average $R$ of the Hall C data. The red square shows the projected precision of the proposed SBS $e^+p$ measurements when combined in a weighted average, plotted arbitrarily at $\epsilon = 0.95$. The pink inverted triangle shows the weighted average of the GEp-2$\gamma$ data for $e^-p$, plotted arbitrarily at $\epsilon = 0.975$.} 
    \label{fig:SBS2gamma}
\end{figure}

It is also desirable to choose a $Q^2$ for which $R_{e^-p}$ is already precisely known. $Q^2 = 2.5$ GeV$^2$ is an obvious choice, being close to the most precise existing measurements~\cite{Puckett:2017flj} in the $Q^2$ region where the discrepancy is significant. At this $Q^2$, it is also possible to measure the $\epsilon$ dependence of $R_{e^+p}$ in a reasonable amount of beam time, as shown in Fig.~\ref{fig:SBS2gamma}. In 60 days' beam time, $R_{e^+p}$ can be measured over a significant range of $\epsilon$ with an absolute statistical uncertainty of 2\% or less. Combining the two measurements in a weighted average would give $\Delta R(e^+p) \approx 0.012$, which is competitive in precision with the best existing electron measurements at this $Q^2$. Another measurement at a meaningfully larger $Q^2 \approx 3.5$ GeV$^2$ would also be attractive, as it would be very close to two existing measurements from the GEp-I~\cite{Punjabi:2005wq} and GEp-II~\cite{Gayou:2001qd} experiments which, however, are significantly less precise. Going significantly higher in $Q^2$ than 3.5 GeV$^2$ would most likely require prohibitive beam time to reach a precision goal of 1-2\%, given the low maximum current for polarized positrons. 

Table~\ref{tab:prop} shows the basic parameters of a plausible program of $R_{e^+p}$ measurements using the SBS GEP apparatus. To estimate the precision of these measurements, elastic $e^+p$ scattering events were generated in a range of angles sufficient to populate the combined acceptance of SBS and ECAL, and tracked through the GEANT4 simulation of the experiment, including the transport of the outgoing proton's spin through the SBS magnetic field. The event rates and the figure of merit for polarimetry were evaluated using the methods described in Ref.~\cite{GEP5_PAC47}, assuming a 200 nA, 60\% polarized $e^+$ beam~\cite{Abb16} on a 40-cm liquid hydrogen target. Since the precision of the existing data at the higher $Q^2$ is only about 4\% (absolute), it would be desirable to include an additional measurement of $e^-p$ scattering in identical kinematics. This could be accomplished in a tiny fraction of the total beam time, as shown in Tab.~\ref{tab:prop}, plus any time that would be required to change CEBAF from positron mode to electron mode and back again. Alternatively, such a measurement could be added to the planned SBS GEP experiment with very little additional beam time, in anticipation of a future comparison to positron measurements.

The systematic uncertainties of the polarization transfer method are typically extremely small. Because both $P_t$ and $P_\ell$ are measured simultaneously in a single kinematic configuration, a number of sources of systematic uncertainty, such as beam polarization and analyzing power, cancel in the ratio $P_t/P_\ell$. The luminosity also doesn't need to be known precisely. Moreover, the $ep$ reaction is self-calibrating with respect to the analyzing power, and the rapid beam helicity reversal cancels the effects of false or instrumental asymmetries in the polarimeter. In previous experiments of this type, a dominant source of systematic error was the calculation of the proton spin precession in focusing magnetic spectrometers with several quadrupole magnets in addition to the main, momentum-analyzing dipole. In the SBS case, the spin precession calculation is much simpler, as the SBS is a single, simple dipole magnet which is non-focusing. It is therefore anticipated that any measurement of $e^+ p$ polarization transfer observables will be statistics-limited in terms of accuracy. In addition, the relatively low luminosity of the proposed $e^+p$ measurements means that the event reconstruction in the SBS detectors will be extremely clean, and far \emph{less} challenging than in the approved high-$Q^2$ measurements from the SBS GEP experiment~\cite{GEP5_PAC47}.

\begin{figure}[ht]
  \begin{center}
    \includegraphics[width=0.98\columnwidth]{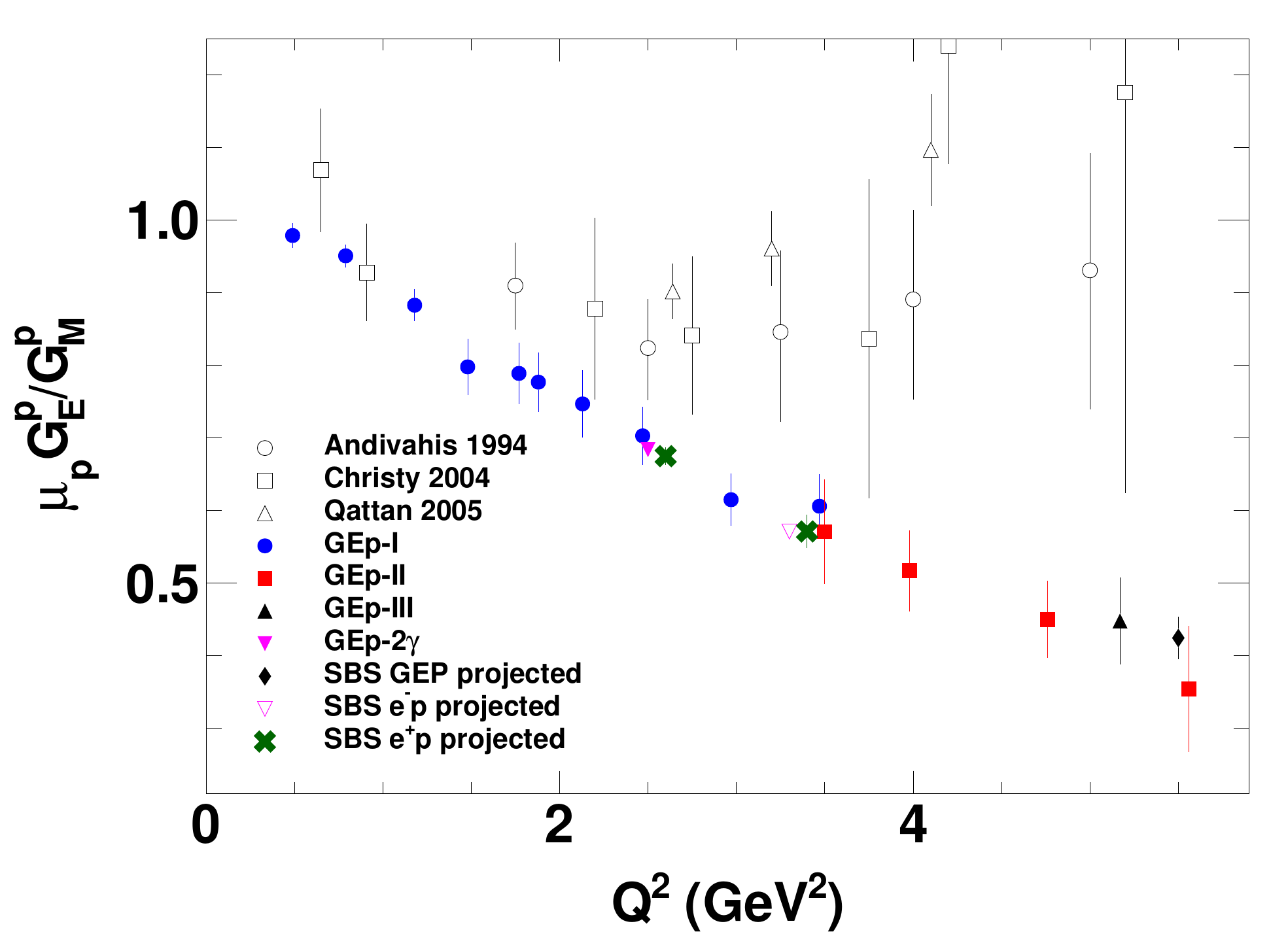}
  \end{center}
  \caption{\label{fig:SBSeplus} Projected results of the proposed future measurements of $R_{e^+p}$ using the polarization transfer method with the SBS GEP apparatus, compared to selected existing data and the projected results of the SBS GEP experiment~\cite{GEP5_PAC47}. Selected Rosenbluth separation data are from Refs.~\cite{Andivahis:1994rq,Qattan:2004ht,Christy:2004rc}. Selected polarization transfer data are from Refs.~\cite{Punjabi:2005wq} (GEp-I), ~\cite{Puckett:2011xg} (GEp-II), and~\cite{Puckett:2017flj} (GEp-III and GEp-2$\gamma$). Projections of future experiments, including the proposed measurements from Tab.~\ref{tab:prop}, are shown at values of $R$ from the global fit described in Ref.~\cite{Puckett:2017flj}, and the projected precision of the SBS result at 2.5 GeV$^2$ is the weighted average of the two measurements at different $\epsilon$ values (see Fig.~\ref{fig:SBS2gamma} and Tab.~\ref{tab:prop}).}
\end{figure}
Figure~\ref{fig:SBSeplus} shows what could be accomplished in the 120-day experiment at ``standard'' CEBAF 1st-pass and 2nd-pass beam energies of 2.2 and 4.4 GeV, under the perhaps somewhat optimistic assumption that a positron beam of 200 nA current and 60\% polarization could be realized at CEBAF. This would be the first measurement of polarization transfer in $e^+p$ scattering, reaching very respectable precision in the $Q^2$ regime where the discrepancy between cross sections and polarization observables is large, and where $R_p$ is falling most rapidly as a function of $Q^2$. Such data would provide important model-independent constraints on hard TPEX amplitudes, toward the goal of finding a conclusive explanation of the discrepancy and a model-independent, data-constrained theoretical prescription for applying hard TPEX corrections to elastic $e^\pm p$ scattering observables. 

\begin{acknowledgements}
This work was supported in part by the US Department of Energy Office of Science, Office of Nuclear Physics, Award ID DE-SC0021200 and in part by the National Science Foundation grant number 2012114
\end{acknowledgements}

%

\bibliographystyle{spphys}       
\bibliography{references}   

\end{document}